\documentclass[aps,prl,twocolumn,groupedaddress,showpacs]{revtex4}
\usepackage{graphicx}

\begin{document}

\title{Anomalous behavior of the Debye temperature in Fe-rich Fe-Cr alloys}

\author{B. F. O. Costa}
\affiliation{CEMDRX Department of Physics, University of Coimbra,
3000-516 Coimbra, Portugal}

\author{J. Cieslak}
\author{S. M. Dubiel}
\email[Corresponding author: ]{dubiel@novell.ftj.agh.edu.pl}
\affiliation{Faculty of Physics and Applied Computer Science,
AGH University of Science and Technology, al. Mickiewicza 30,
30-059 Krakow, Poland}

\date{\today}

\begin{abstract}
Debye temperature, $\Theta_D$, of Fe-rich Fe$_{100-x}$Cr$_x$ disordered alloys
with $0\le x \le 22.3$ was determined from the temperature dependence of the
central shift of M\"ossbauer spectra recorded in the temperature range of 60
-- 300 K.  Its compositional dependence shows a maximum at $x \approx 5$ with
a relative increase of $\sim 30$\% compared to a pure iron.  The composition
at which the effect occurs correlates well with that at which several other
quantities, e.  g.  the Curie temperature and the spin-wave stiffness
coefficient, $D_0$, show their maxima, but the enhancement of $\Theta_D$ is
significantly greater and comparable with the enhancement of the hyperfine field (spin-density
of itinerant $s$-like electrons) in the studied system.  The results suggest
that the electron-phonon interaction is important in this alloy system.
\end{abstract}

\pacs{63.20.-e, 75.50.Bb, 76.80.+y}

\maketitle

Fe$_{100-x}$Cr$_x$ alloys are both of scientific and technological interests.
The former follows, among other, from the fact that they can be regarded as a
model system for studying various magnetic properties and testing appropriate
theoretical models.  The latter is related to the fact that the alloys form a
matrix for a production of chromium steels that, due to their excellent
properties, find a wide application in industry \cite{Hishinuma02}.  For example, the steels
containing 2--20~at\% Cr are regarded as good candidates for the design of
structural components in advanced nuclear energy installations such as
Generation IV and fusion reactors.  In that range of composition, the alloys
show an anomalous behaviour in that several physical quantities exhibit
extreme values.  However, their position and the relative value depend on the
quantity.  For example, the Curie temperature, $T_C$, has its maximum value at
$x \approx 5$~at\%, which is paralleled by a maximum in the neutron value of
the spin-wave stiffness coefficient, $D_0$.  However, the relative increase in
$T_C$ from the pure Fe is only $\sim$1\% \cite{Adcock31} compared with the $\sim$10\% effect
in the $D_0$ value \cite{Lowde65}.  An enhancement was also found in the hyperfine (hf)
field as measured at $^{57}$Fe nuclei \cite{Dubiel76} as well as at $^{119}$Sn nuclei \cite{Dubiel80}.
In both cases the maximum was for $x \approx 10$~at\% with the relative
increase of $\sim$4\% for the former and $\sim$15\% for the latter probe
nuclei.  Hyperfine field is usually positively correlated with the magnetic
moment, $\mu$, hence in the light of the above mentioned results, one should
expect a paralleled behavior for the latter.  Indeed, an increased value of
$\mu$ localized at Fe site was revealed from neutron diffraction experiments
\cite{Aldred76,Lander71}.  In this case the maximum occurs at $x \approx 20$~at\% and its
relative enhancement is equal to $\sim$7\%.  Increment of $\mu$ was predicted
theoretically to exist at $x \approx 15$~at\% with the relative effect of
$\sim$17\% \cite{Frollani75},as well as at $x \approx 5$~at\% with the relative effect of
$\sim$3\%\cite{Olsson06}, and, that of $T_C$ at $x\approx 15$~at\% and with the effect of
$\sim$25\% \cite{Kakehashi87}.  The phenomenon in the latter was explained by the enhancement
of Fe-Fe exchange coupling due to the alloying effect.  The importance of
magnetism in the understanding of these alloys, in general, and Fe-rich ones,
in particular, also seems to be crucial in the light of recent theoretical
calculations that predict a negative sign of the heat of formation of Fe-Cr
alloys with Cr content less than 10--12~at\% \cite{Olsson06,Klaver06} as well as those obtained
with ab initio calculations combined with synergic synchrotron x-ray
absorption experiments showing an anomaly at $x \approx 13$~at\% \cite{Froideval07}.  Also a
drastic decrease in the corrosion rate with chromium content increase occurs
within a concentration range of 9-13 at \% \cite{Wranglen85}, but its reason is not fully
understood.  Recent calculations based on first-principles quantum-mechanical
theory shed some light on the issue.  In particular, they demonstrate that
within this concentration range there is a transition between two surface
regimes; for bulk Cr content greater than $\sim$10~at\% the Cr-rich surfaces
become favorable while for a lower concentration the Fe-rich ones prevail
\cite{Ropo07}.  According to the authors of Ref.  \onlinecite{Ropo07}, they are related with two
competing magnetic effects:  the magnetically induced immiscibility gap in
bulk Fe-Cr alloys and the stability of magnetic surfaces.  In addition, the
authors show that other theoretical bulk and surface properties of Fe-rich
ferromagnetic Fe-Cr alloys have their minima viz.  effective chemical
potential at $x \approx 15$~at\%, and the mixing enthalpy at $x \approx 4$~at\%
\cite{Ropo07}.  The latter concentration coincides well with that at which a
change in the sulphidation preference takes place \cite{Cieslak93}.

For the complete, or at least, a better understanding of all theses effects
and their relationship, further theoretical and experimental studies are,
however, necessary.  Towards the latter end, we have carried out measurements
of the Debye temperature, $\Theta_D$, for $0 \le x \le 22.3$ in order to verify
whether or not this quantity related to the dynamics of the lattice exhibits
an enhancement of its value.  Though it is widely believed that the possible
effect of the electron-phonon interaction on the magnetism of metallic system
is of no significant importance \cite{Herring66}, because the modification of the exchange
interaction or magnetization caused by the interaction should be of the order
of $10^{-2}$.  However, calculations by Kim \cite{Kim82,Kim88} suggest that in an
itinerant-electron ferromagnet the electron-phonon interaction might be
important.  Including exchange interaction between electrons enhances the
effect by a factor of $\sim$10 to $\sim$100 \cite{Kim82,Kim88}.  If these calculations are
correct, then a correlation between magnetic and dynamic properties should be
experimentally observed.  Although in the literature the are already some data
on the Debye temperature in the Fe-Cr system, they are not conclusive in that
respect as they were measured in a very narrow and low temperature range viz.
1.4 -- 4.2 K \cite{Cheng60}, and, in the concentration range of interest, there are
only three data points.
A more detailed study of the issue was also prompted by an anomalous behavior
of $\Theta_D$ found for the Cr-rich samples, and, in particular, by the smallest value of $\Theta_D$
for the most Cr-rich alloy.

\begin{figure}[htb]
\includegraphics[width=.42\textwidth]{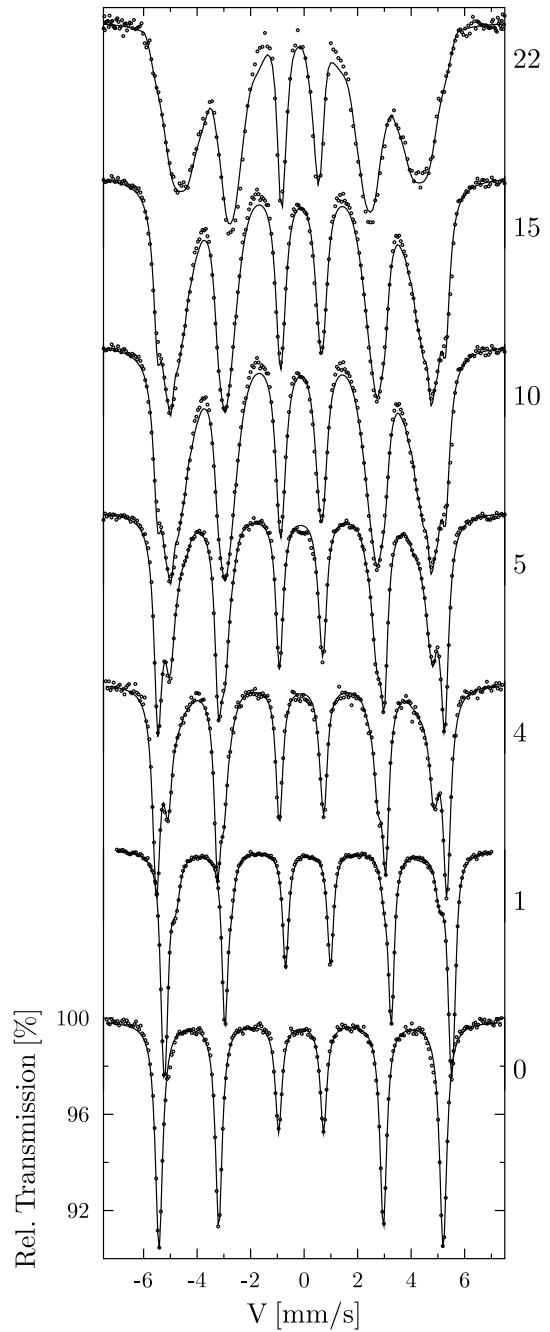}
\caption{Room temperature M\"ossbauer spectra recorded on Fe$_{100-x}$Cr$_x$ samples for
various $x$-values (0) $x$ = 0, (1) $x$ = 1.3, (4) $x$ = 3.9, (5) $x$ = 4.85, (10)
$x$ = 10.25 and (22) $x$ = 22.3. The solid lines are the best-fit to experimental data. }
\label{fig1}
\end{figure}

\begin{figure}[htb]
\includegraphics[width=.42\textwidth]{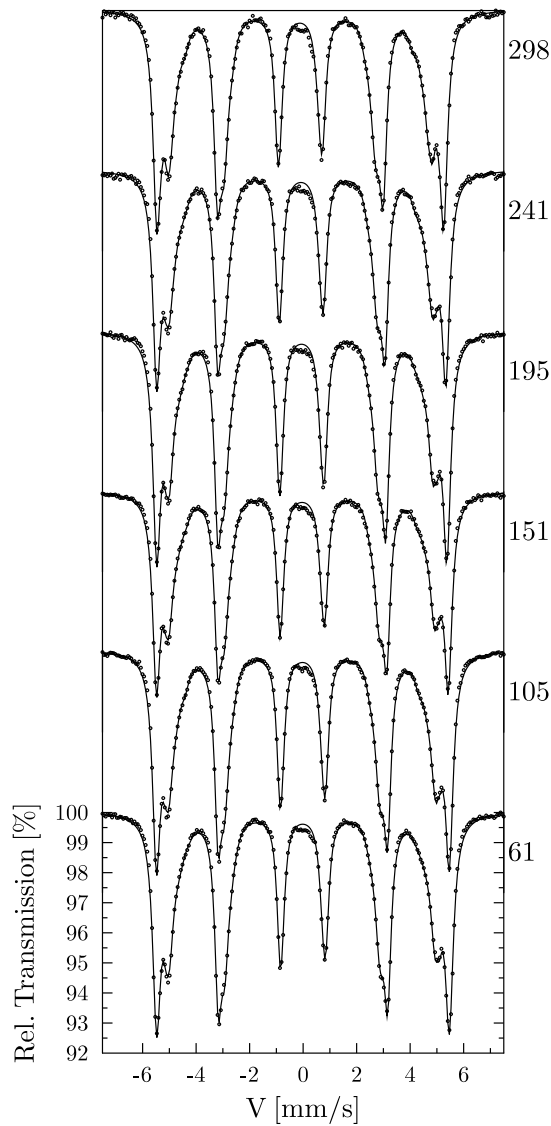}
\caption{M\"ossbauer spectra recorded on Fe$_{95.15}$Cr$_{4.85}$ sample at different temperatures
(in Kelvin) shown. Solid lines represent the best-fits to the experimental spectra.  }
\label{fig2}
\end{figure}

For the present study previous samples were used \cite{Dubiel76,Dubiel81}.  The Debye
temperature was measured by means of the M\"ossbauer spectroscopy.  For that
purpose a series of M\"ossbauer spectra was recorded in a transmission
geometry for each sample in the temperature range of 60 -- 300 K using a
standard spectrometer and a $^{57}$Co/Rh source of 14.4 keV gamma rays.
Temperature of the samples which were kept in a cryostat were stabilized with
an accuracy of $\sim$0.2 K during measurements.  Examples of the recorded spectra are shown in
Figs.  1 and 2.  They were fitted in two ways, I and II, to get an average
value of the central shift, $<CS>$, which is the quantity of merit for
determining $\Theta_D$.  In way I, each spectrum was fitted assuming it
consists of a number of six-line patterns subspectra, each of them
corresponding to a particular atomic configuration around the probe $^{57}$Fe
nucleis, $(m,n)$, where $m$ is a number of Cr atoms in the first neighbor
shell (NN), and $n$ is a number of Cr atoms in the next neighbor shell (NNN).
It was further assumed that the effect of neighboring Cr atoms on spectral
parameters (hyperfine field, and central shift) was additive.  Using this
procedure, which is described elsewhere in detail \cite{Dubiel76,Dubiel80,Dubiel81}, the average
central shift, $<CS>$, could have been calculated.  In way II, each spectrum
was fitted in terms of the hyperfine field distribution method \cite{LeCaer79}.
Following the experimental result \cite{Dubiel81}, a linear correlation between the
hf.  field and the isomer shift was assumed in the fitting procedure.  The
Debye temperature was determined from the temperature dependence of $<CS>$:

\begin{equation}
<CS(T)> = IS(T) + SODS(T)  \label{equ1}
\end{equation}

where $IS(T)$ is the isomer shift and it is related to the charge density at the
probe nucleus and has a weak temperature dependence \cite{Willigeroth84}, so it is usually
approximated by a constant term, $IS(0)$, which is eventually composition dependent.
$SODS$ is the
so-called second-order Doppler shift which shows a strong temperature
dependence.  Assuming the whole temperature dependence of $<CS>$ goes via
$SODS$ term and using the Debye model for the phonon spectrum one arrives at
the following formula:

\begin{equation}
CS(T) =IS(0)-{3kT \over 2mc}\left[{3\Theta_D \over 8T}+ 3\left({T \over \Theta_D}\right)^3
\int^{\Theta_D \over T}_0 {x^3 \over e^x-1} dx \right]    \label{equ2}
\end{equation}

\begin{figure}[htb]
\includegraphics[width=.45\textwidth]{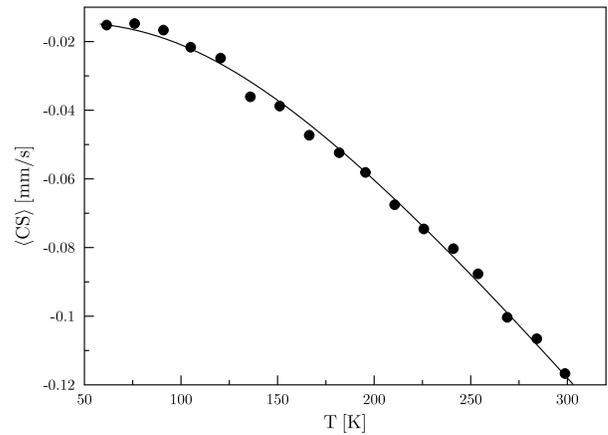}
\caption{Dependence of the average central shift, $<CS>$, on temperature for
Fe$_{95.15}$Cr$_{4.85}$ alloy. The solid line represents the best-fit to the
experimental data in terms of equation (\ref{equ2}).}
\label{fig3}
\end{figure}

where $m$ is the mass of the $^{57}$Fe nucleus, $k$ is the Boltzmann constant,
$c$ is the velocity of light.  Fitting equation (\ref{equ2}) to the $<CS(T)>$ - values
determined in the ways I and II (whose typical temperature behavior is
illustrated in Fig.  3), enabled determination of the $\Theta_D$ - values.  It
has turned out that within the error limit they agree well with each other,
hence, for final considerations, the average, $<CS> = 0.5 (<CS>_I + <CS>_{II})$,
between each corresponding pair has been taken.  Its normalized ($\Theta_D$ =
429 K was found for a pure Fe - see the Table \ref{table1}) dependence on Cr content, $x$, is shown in Fig.
4, where, for comparison, normalized values of other physical quantities that
have maximum in the their value are shown.

\begin{table}[hb] 
\caption{\label{table1} Determined $\Theta_D$ values of Fe-rich Fe$_{100-x}$Cr$_x$ disordered alloys.
$\Delta\Theta_D$ stands for the $\Theta_D$-error.
Values of $IS(0)$ relative to Co/Rh source are also displayed.}
\begin{tabular}{|lr|r|r|r|r|r|r|r|r|} \hline
$x$             &       &    0.0 &   1.2&  3.9 &  4.8 &  8.6 & 10.2 & 15.0 & 22.3           \\ \hline
$\Theta_D$      & [K]   &    429 &   493&  557 &  548 &  569 &  519 &  524 &  463           \\ \hline
$\Delta\Theta_D$& [K]   &     14 &    11&   15 &   11 &   18 &   18 &   33 &   29           \\ \hline
$IS(0)$         &[mm/s] &   0.006&-0.024&-0.030&-0.014&-0.027&-0.020&-0.033&-0.031          \\ \hline
\end{tabular}
\end{table}

From the $IS(0)$-data shown in Table I it is clear that $IS(0)$ hardly depends on $x$, hence
the anomaly observed in $\Theta_D(x)$ has its full origin in $SODS$.
The concentration at which the
maximum occurs is similar or very close to that at which several other
quantities or phenomena show anomalies, and, in particular, (a) the Curie
temperature \cite{Adcock31} and (b) the neutron value of the spin-wave stiffness
coefficient, $D_0$ \cite{Lowde65}.  The increase of the former can be explained in terms
of the increase of the magnetic bonding between Fe atoms \cite{Kakehashi87}.  However,
no quantitative agreement between the enhancement factor of $T_c$ and $Theta_D$ exists.
That
of the latter is significantly greater and it is rather close to the
enhancement of the $^{119}$Sn-site hyperfine field, $H(0,0)$, but not to the
one of the $^{57}$Fe-site hyperfine field, $H(0,0)$.  As the former
mostly originates from the polarization of the conduction (itinerant)
electrons and the latter from the polarization of the core (localized)
electrons, one is tempted to conclude that for the enhancement of $\Theta_D$
revealed in this study the itinerant electros are predominantly responsible.

\begin{figure}[htb]
\includegraphics[width=.45\textwidth]{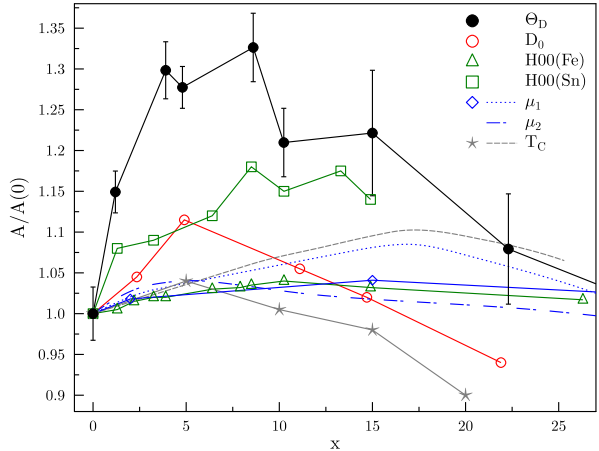}
\caption{(Color online) Normalized values of various physical quantities, $A/A(0)$, as measured
(presented with various symbols connected by lines to guide the eye) or calculated
(for the magnetic moment $\mu_1$ illustrated by a dotted line \cite{Frollani75}
and for $\mu_2$ by dashed-dotted line \cite{Olsson06} as well as the Curie point by a
dashed line \cite{Kakehashi87})
for Fe$_{100-x}$Cr$_x$ alloys that show a maximum in the concentration range
of $0 \le x \le 26$.}
\label{fig4}
\end{figure}

In other words, the change of the Debye temperature of Fe-Cr alloys in the
investigated range of their composition reflects underlying change in the
polarization of the conduction electrons.  This in turn, can be taken as
evidence that in the investigated alloy system there is a coupling between
spin-polarization and the phonon spectrum.  A lack of a quantitative agreement
as measured in terms of the position of the maximum and the amplitude of the
enhancement means that the coupling is not linear.

\begin{acknowledgments}
The results presented in this paper were obtained in the
frame of the bilateral Polish-Portuguese project 2007/2008.
\end{acknowledgments}

\end{document}